# Characterization of Adiabatic Quantum-Flux-Parametrons in the MIT LL SFQ5ee+ Process

Sergey K. Tolpygo, *Senior Member, IEEE,* Evan B. Golden, Christopher L. Ayala, *Senior Member, IEEE,* Lieze Schindler, *Member, IEEE,* Michael A. Johnston, *Member, IEEE,* Neel Parmar, and Nobuyuki Yoshikawa, *Fellow, IEEE*

*Abstract*—**Adiabatic quantum-flux-parametron (AQFP) superconductor logic is a proven energy-efficient digital technology for various applications. To address the scalability challenges of this technology, we investigated AQFP shift registers with the AQFP footprint area reduced by 25% with respect to prior work and with the more than 2× denser overall circuit designs obtained by eliminating the previously used free space between the AQFPs. We also investigated AQFP cells with different designs of flux trapping moats in the superconducting ground plane as well as compact AQFP cells that took advantage of the smaller feature sizes available in the new fabrication process, SFQ5ee+, at MIT Lincoln Laboratory (MIT LL). This new process features nine planarized Nb layers with a 0.25 μm minimum linewidth.**

**The fabricated circuits were tested in a liquid He immersion probe and a commercial closed-cycle cryocooler using a controlled cooling rate through the superconducting critical temperature, $T_c$. Using multiple thermal cycles, we investigated flux trapping in the dense AQFP shift registers as well as in the registers using the old (sparse) AQFP designs at two levels of the residual magnetic field, about 0.53 μT and about 1.2 μT. The sparse designs demonstrated 95% to almost 100% probability of operation after the cooldown and very wide operation margins, although the flux trapping probability was increasing with circuit complexities. The margins were similarly wide in the newer dense designs, but flux trapping probability that rendered the registers nonoperational was significantly, in an order of magnitude, higher in these denser circuits and was also very sensitive to the moats' shape and location.**

**Our findings indicate that AQFP circuits are amendable to increasing the scale of integration and further densification, but a careful moat design and optimization are required to reduce flux trapping effects in the dense AQFP circuits.**

## I. INTRODUCTION

THE electricity demands to support internet of things (IoT), Big Data, and social media have been rising rapidly, and it is unclear whether conventional CMOS technology can continue to serve as the technological platform to support these services in the long term. Adiabatic quantum-flux-parametron (AQFP) superconductor logic [1] is an alternative technology that can potentially fulfill this role. When using a 100 μA/μm$^2$ unshunted Nb/Al-AlO$_x$/Nb Josephson junction (JJ) fabrication process available today [2], AQFPs can operate with switching energies approaching a zeptojoule per JJ at GHz clock rates [3]. An important milestone for this technology was a successful demonstration of a 4-bit microprocessor called MANA with over 20k JJs [4]. It helped identify new research areas to improve the scalability of AQFP technology [5]. Such research directions include reducing the area footprint of the AQFPs for more dense designs, improving the robustness of the AQFP cells to unwanted flux trapping, and identifying the origin of AQFP input offsets which may cause increasing the bit error rate.

Using a new fabrication process for superconductor electronics developed at MIT Lincoln Laboratory (MIT LL), we reduced the footprint of the AQFP buffer cells by 25%, from 15 μm x 20 μm [6] in the original SFQ5ee process [2] to 15 μm x 15 μm in the new SFQ5ee+ process, and increased the overall density of the multi-stage AQFP shift register by more than a factor 2×. To investigate flux trapping in these dense AQFP circuits we designed shift registers with different configuration of flux trapping moats. The new fabrication process, SFQ5ee+, was presented in [5] and is briefly described in Sec. II along with the new circuit designs and their comparison with the sparse designs used in [6]. Electrical testing results are given in Sec. III. They show functional AQFP shift registers comprising up to 69 stages and very wide operating margins approaching ±50% with respect to the ac excitation flux. However, we have found a significantly higher flux trapping probability in denser

This material is based upon work supported in part under Air Force Contract No. FA8702-15-D-0001 and by the Grant-in-Aid for Scientific Research (S) No. 19H05614 and (C) No. 21K04191 from the Japan Society for the Promotion of Science (JSPS). *(Corresponding author: Sergey K. Tolpygo, e-mail: sergey.tolpygo@ll.mit.edu).*

S.K. Tolpygo, E.B. Golden, and N. Parmar are with the Massachusetts Institute of Technology, Lexington, MA 02421, USA (e-mails: sergey.tolpygo@ll.mit.edu, evan.golden@ll.mit.edu, and neel.parmar@ll.mit.edu).

C.L. Ayala, L. Schindler, M.A. Johnston, and N. Yoshikawa are with the Institute of Advanced Sciences, Yokohama National University, Hodogaya, Yokohama 240-8501, Japan (emails: ayala-christopher-pz@ynu.ac.jp, liezeschindler@gmail.com, mjohnston@ieee.org, nyoshi@ynu.ac.jp). L. Schindler is also with SUN Magnetics and Stellenbosch University, Stellenbosch, 7600 South Africa.

AQFP circuits than in the sparse designs used in [6], which was also very sensitive to the configuration of moats in the AQFP ground plane. Potential reasons for this are discussed in Sec. IV.

## II. AQFP Design, Fabrication Process, And Testing

A schematic cross section of the SFQ5ee+ process is shown in Fig. 1 [5]. The main differences from the traditional SFQ5ee process [2] are: a) an additional Nb layer, ninth Nb layer, above the layer of Josephson junctions, layer M8; b) reduced minimum linewidth to 0.25 μm; c) allowing stacked vias.

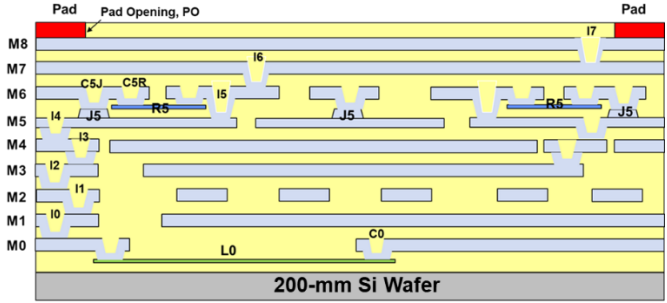

**Fig. 1.** Schematic cross section of the SFQ5ee+ fabrication process [5], showing Nb layers M0-M8, kinetic inductor layer L0, shunt resistors R5, and Josephson junctions J5. Stack-staggered vias I0-I4 are also shown; vias to the kinetic inductors, junctions, and shunt resistors are, respectively C0, C5J, and C5R.

### A. AQFP Cell Design

Reducing cell size and improving circuit density is a pressing requirement in the development of superconducting circuits. One of the goals of this work was to reduce the size of the AQFP cell within the constraints of the new SFQ5ee+ process.

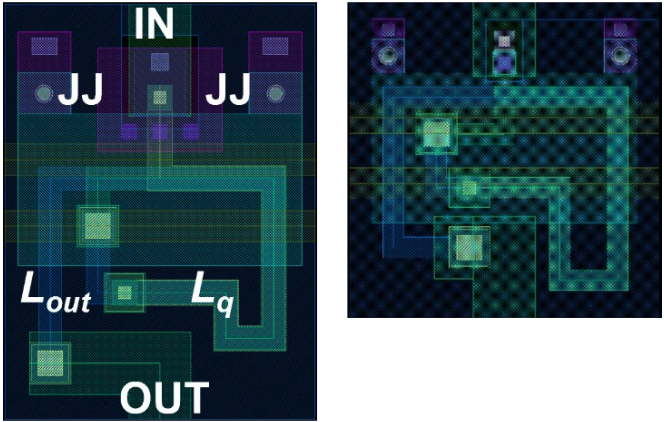

**Fig. 2.** Layouts of the AQFP buffer in the SFQ5ee process (left) used in [6], referred hereafter as the reference design, and in the SFQ5ee+ process (right). The buffer cell was 15 μm x 20 μm [6] and became 15 μm x 15 μm after redesign done in the work. Note the reduction in the linewidth of the AQFP inductors $L_q$ and $L_{out}$; and reduction in the size of all vias; see notations in [1], [5]. The ground plane is Nb layer M4 in Fig. 1. Inductors $L_q$ and $L_{out}$ are formed using layers M1-M3 below the ground plane. AQFP dc flux bias and ac excitation are applied using microstrip transmission lines (horizontal lines in the picture) on layer M7. The AQFP dc-SQUID inductor is formed between layers M6 and M4.

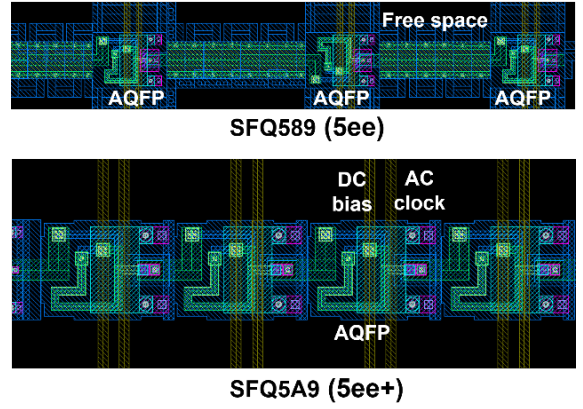

**Fig. 3.** Fragments of the layouts of a sparse AQFP buffer chain in the SFQ5ee process (top) used in [6] and of the more compact chain in the SFQ5ee+ process, using the same reference buffer, run SFQ5A9, (bottom). The pictures have different scale, but the circuit density comparison can be done using the AQFP dimensions given in Fig. 2.

The compact cells were developed by using proven AQFP cells [6] as a reference and starting point; see Fig. 2. The features of the reference design were minimized where possible while keeping the functionality and excitation parameters of the AQFP cell identical. This required the ac excitation and dc flux bias lines to be left unchanged whereas JJs, via stack sizes and locations, as well as coupling inductors and routing were adjusted in line with the SFQ5ee+ constraints. This allowed the compact cells to be integrated with reference cells and ensured that testing could be carried out using known operation bias points. The cell design process was iterative - InductEx [7] was used to extract the relevant parameters and modifications were made to ensure that design values were achieved.

Using this process, we designed the following AQFP test circuits:

1. AQFP cells with different flux trapping moat designs and configurations.
2. 69-stage AQFP shift registers to evaluate functional performance and yield in highly dense structures.
3. Asymmetric AQFPs to help identify a model-to-hardware correlation of the input offsets in AQFPs.
4. Compact AQFP cells that take advantage of the smaller feature sizes of the SFQ5ee+ process.

### B. AQFP Layouts

The modifications resulted in a size reduction from 15 μm x 20 μm to 15 μm x15 μm for an AQFP buffer cell. Fig. 2 shows differences in the layouts of the AQFP buffers fabricated in this work (right panel) in the SFQ5ee+ process, run SFQ5A9, and those fabricated in the SFQ5ee process, run SFQ589 from the previous work [6]. Note that the cell width remained unchanged due to the length requirements of the excitation inductors (two horizontal wires in Fig. 2). We expect that further reduction in cell size may be achieved if the excitation parameters are re-designed with the SFQ5ee+ process in mind and through the elimination of the on-chip distribution of the dc offset required in implementation of the 4-phase excitation network [8] by providing dc offsets off-chip as done in [9].

Fig. 3 shows a fragment of the AQFP buffer chain used in [6]

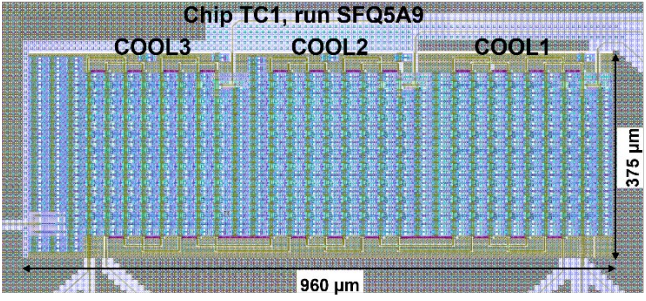

**Fig. 4**. Layouts of test chip, TC1, containing three 69-stage AQFP shift registers, COOL1, COOL2, and COOL3 differing by the design of flux trapping moats in the M4 ground plane. Only COOL1 and COOL2 shift registers were tested in this work. These dense chains use the reference AQFP buffer design.

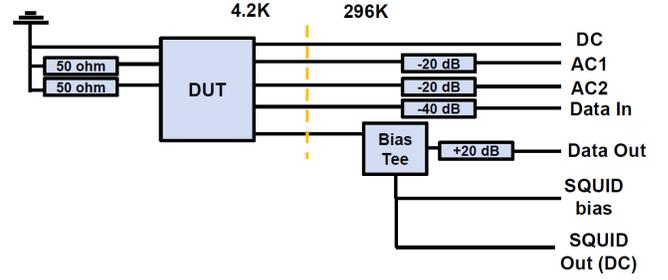

**Fig. 6**. Block diagram of a test setup used for testing AQFP circuits. A four-channel arbitrary waveform generator provided ac excitation signals AC1 and AC2 with 90 degrees phase shift via 50-ohm coaxial lines. The data were acquired using a 4-channel digital oscilloscope and a low noise amplifier, amplifying signals from the output dc-SQUID. Yokogawa current sources were used to supply dc flux bias (DC) and the output SQUID bias.

(top panel) and of a compact buffer chain designed in the SFQ5ee+ process (bottom panel). In the previous design [6], AQFPs in the chain were separated by a distance twice as large as the size of the AQFP itself and the data were transferred using a long microstrip inductor $L_{out}$. In the compact chain design, the AQFP cells are abutted to each other using minimum spacing between the cells. This more than doubled the circuit density.

69-stage AQFP shift registers using the reference AQFP buffer shown in Fig. 2 were designed without free space between the cells as shown in the bottom panel of Fig. 3. Designs with differing ground plane moat structures were arranged into a chip shown in Fig. 4.

### C. Flux Trapping Moats in AQFP Layouts

Chips designed for the SFQ5ee+ process run SFQ5A9 used trial fill and moat structures. The moat design of two shift registers cells, COOL1 and COOL2, is shown in Fig. 5. A single ground plane M4 was used. Moats can be seen as white square, rectangular, and L-shaped openings repeated around each AQFP cell in the registers. The only difference in the designs of COOL1 and COOL2 is that L-shaped moats in COOL2 were removed (not defined in the M4 layer), whereas they are identical in all other respects. This allowed for testing of the effectiveness of the L-shaped moats which were originally placed very close to the AQFP output transformer.

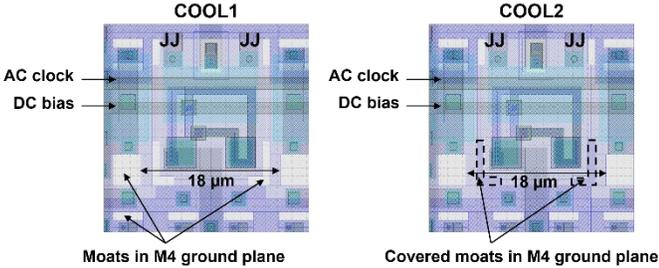

**Fig. 5**. Layouts of flux trapping moats in 69-stage AQFP shift registers COOL1 and COOL2. Moats can be seen as white color openings in the M4 ground plane background. The only difference between the moats in registers COOL1 and COOL2 is the absence of L-shaped moats near the AQFP output transformer in COOL2 (shown as covered moats).

### D. Testing

The test circuits were fabricated on three 5 mm x 5 mm chips,

via fabrication run SFQ5A9 and tested at MIT LL. Two test setups were used: a) a liquid helium immersion probe with two nested mu-metal shields and a residual magnetic field of about 0.53 µT; b) a commercial pulse-tube cryocooler-based setup with a single mu-metal shield with a residual magnetic field of about 1.2 µT. For comparison, we used sparse AQFP circuits designed for the standard SFQ5ee process and fabricated in a prior run (SFQ589) as a reference. Their successful operation was previously demonstrated in [6]. Low speed testing was done using the newest version of a 128-channel automated setup Octopux [10], [11]. High speed testing was done in the LHe immersion probe, using a setup shown in Fig. 6, which is similar to the one described in [6].

### III. TEST RESULTS

### A. Sparse AQFPs with the Reference Design [6]

The typical results of low-speed testing of the AQFPs with the reference sparse design [6] are shown in Fig. 7. The purpose of this testing was to verify the test setup, reproduce the test results in [6], and study flux trapping in the AQFPs with the sparse design fabricated in the SFQ5ee process. Multiple circuits fabricated on the chip were tested giving nearly identical operating margins:

a) ac clock amplitude: 450 µA – 1050 µA (±40%)
b) dc flux bias: 1.00 mA – 1.95 mA (±32%)
c) readout SQUID bias: 155 µA – 185 µA

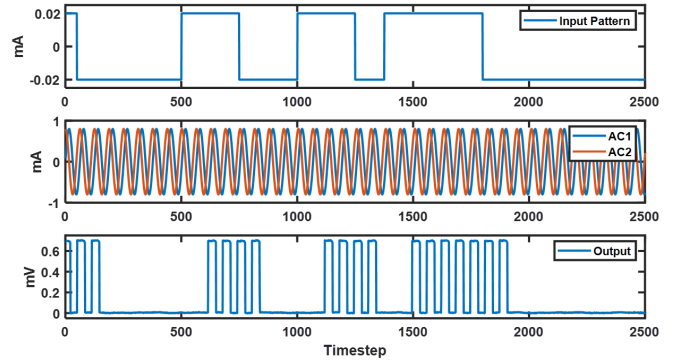

**Fig. 7**. Testing of a sparse AQFP buffer chain fabricated in the SFQ5ee process run SFQ589 and described in [6]. The panels from top to bottom show the input data, the ac excitation signals AC1 and AC2, and the output SQUID voltage.

Functionality of all devices on the chip and their propensity to flux trapping was tested in the cryocooler-based setup over many thermal cycles using a cooling rate through $T_c$ of 0.1 K/min (1.7 mK/s). The results are given in Table I. We note the increase of flux trapping probability with increasing circuit complexity, e.g., the XOR gate.

### B. 69-stage AQFP Buffer Chains COOL1 and COOL2

The typical results of low-speed testing of the redesigned AQFP buffer chains in the SFQ5ee+ process, run SFQ5A9 chip TC1 in Figs. 4 and 5, are shown in Fig. 8.

### TABLE I
#### FLUX TRAPPING IN AQFP CELLS WITH SPARSE DESIGN*

| Device name | No. of thermal cycles | No. of fully operational cases | Probability of operation without flux trapping (%) |
|---|---|---|---|
| AQFP Buffer | 57 | 57 | 100 |
| Inverter | 161 | 161 | 100 |
| Majority 3 (MAJ3) | 37 | 37 | 100 |
| AND | 57 | 55 | 96 |
| XOR | 57 | 50 | 88 |

*The circuits were fabricated in the SFQ5ee process and tested in the cryocooler-based setup with residual field of about 1.2 µT.

Similarly, the propensity to flux trapping was investigated in the cryocooler-based setup with residual magnetic field of 1.2 µT. The same cooling rate of 0.1 K/min was used as for the reference design in III.A. The results are shown in Table II.

### TABLE II
#### FLUX TRAPPING IN DENSE AQFP CIRCUITS IN SFQ5EE+

| Device name | No. of thermal cycles | No. of fully operational cases | Probability of operation without flux trapping (%) |
|---|---|---|---|
| 72-stage buffer chain COOL1, cryocooler | 57 | 2 | 3.5 |
| 72-stage buffer chain COOL2, cryocooler | 57 | 7 | 12.3 |
| Compact buffer chain, cryocooler | 380 | 0 | 0 |
| Compact buffer chain, LHe probe* | 5 | 1 | 20% |

*LHe immersion probe with residual magnetic field of about 0.53 µT.

### C. Compact buffer chain

The densest AQFP circuits designed in the SFQ5ee+ process, were chains of the compact buffers shown in Fig. 2, right panel. An optical image of a fragment of the fabricated chain is shown in Fig. 9. In this top view, the bottom ground plane, dc and ac excitation lines, and the AQFP inductor, $(L_1 + L_2)$ in notations of [1], [5] are clearly visible. The output inductors $L_q$ and $L_{out}$ are under the ground plane M4 and are not visible in the picture. The main feature of this compact design is that the compact AQFP does not have its own moats in the cell, and the moats are added to the "dummy" fill cells between the AQFPs on the same row (along the same ac excitation line).

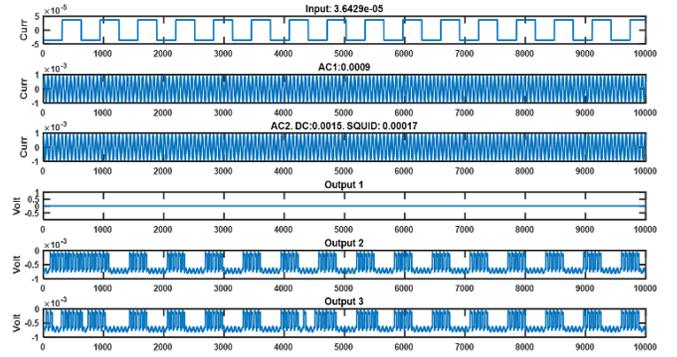

**Fig. 8.** Low-speed testing results of 69-stage buffer chains COOL1 and COOL2 fabricated in the SFQ5ee+ process run SFQ5A9 TC1. The panels from top to bottom show the input data, the ac excitation signals AC1 and AC2, and the output SQUID voltage of COOL2 (Output 2) and COOL1 (Output 3) chains.

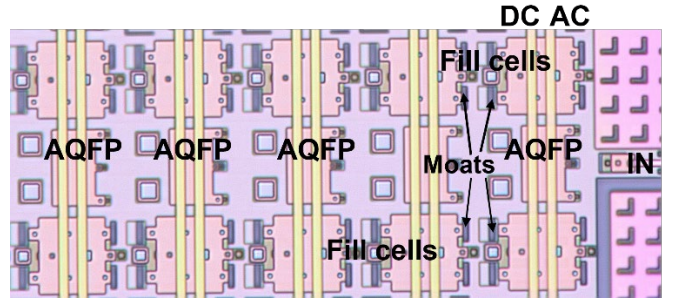

**Fig. 9.** Optical image of a fragment of a 69-stage chain of compact buffers (Fig. 2 right panel) fabricated in the SFQ5ee+ process run SFQ5A9, TC2, test circuit COMP02. Moats in the ground plane of the dummy fill cells are clearly visible as dark rectangles. The compact AQFP buffers do not have their own moats. Brighter vertical lines are dc flux bias and ac excitation microstrips.

The 69-bit chain of compact buffers was tested in LHe immersion probe, using the test setup in Fig. 6. The typical results are shown in Fig. 10. However, flux trapping in the compact buffer chains was the most pronounced of all circuits designed in the SFQ5ee+ process, shown in the last two columns of the Table II. The circuit was not operational in the cryocooler-based test setup despite 380 thermal cycles, and was operational only once out of five cooldowns in the LHe probe with ~2x lower residual magnetic field. The latter indicates that there are no design errors in the circuit, and that the problem is the very high sensitivity of this compact AQFP design to flux trapping.

### IV. DISCUSSION AND CONCLUSION

Our results clearly show that AQFP circuits can be scaled up and significantly densified beyond the levels achieved previously in [6]. However, flux trapping in dense circuits becomes a problem. Since flux trapping events in different cells of a circuit are likely completely independent, the probability of a correct operation (no flux trapping) of an $N$-cell circuit is $P = p^N$, where $p$ is the probability of no flux trapping in the unit cell, which depends on the residual magnetic field and the cell design. This probability can be estimated from the results of our thermal cycling experiment: $p = \exp{(N^{-1}\ln P)}$, using the data in Table II. We get $P \approx 95\%$ and 97% for the COOL1 and COOL2 chains, respectively. These probabilities are about the same as those found for AQFP simple cells with the reference design [6] and also agree with the limited thermal cycling results presented in [6].

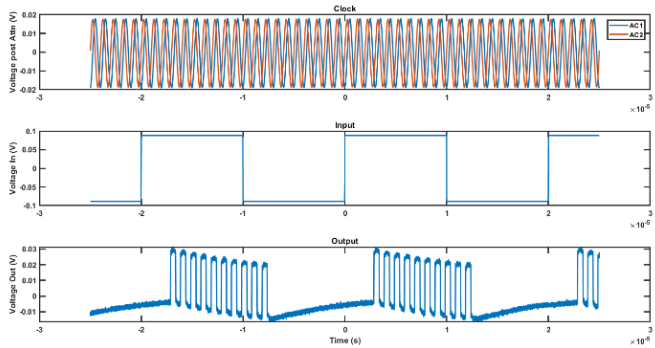

**Fig. 10.** Typical testing results for a chain of compact buffers fabricated in SFQ5ee+ process run SFQ5A9, CH2, test circuit COMP02, shown in Fig. 9. From top to bottom are shown AC1 and AC2 clocks, input data pattern, and the chain output. The test parameters used are as follows: SQUID bias 162 μA; dc flux bias 1.46 mA; ac excitation (clock) amplitude 349 μA; Input data amplitude 38 μA, split in 4 devices.

The width, $w$, of the main ground plane track between the moats in the COOL1 and COOL2 chains is about 18 μm as shown in Fig. 5. The critical perpendicular magnetic field, the flux expulsions field, for the strip of this width can be estimated using, e.g., $B_{c1} = \pi\Phi_0/4w^2$ [12] or $1.65\Phi_0/w^2$ [13], or other more complicated expressions depending on the film shape; see, e.g. [14]. This gives $B_{c1} \approx 5$ μT, a much larger field than the residual magnetic field in any of the setups used for testing. Therefore, Abrikosov vortices are supposed to be unstable in the Nb ground plane film at 4.2 K unless they are strongly pinned by some defects. This analysis does not take into account, however, a parallel component of the magnetic field which may cause flux trapping in the loops of AQFP inductors $L_q$ and $L_{out}$. In this case, the trapped fluxons will create offsets to the inputs of the AQFPs, interfering with the data.

The SFQ5A9 design (separate from the compact cell design) used trial fill and moat structures. Similar structures have been used in other designs but these structures still require thorough evaluation. The combination of compact cells, trial fill and moat structures may have produced unreliable circuit operation. A better approach for future compact circuits would be to fabricate test circuits with the standard fill and moat structures, e.g., similar to [6], along with the trial structures for more thorough testing. InductEx [7] provides functionality to perform flux trapping analysis which could assist in improving fill and moat structure design. InductEx allows to simulate couplings between the fluxons in the moats and all inductors in the cell. These couplings can then be used in the refined netlist to simulate operating margins of the cell using SPICE, e.g., JoSIM [15], and iteratively refine the configuration and positions of the moats. Examples of such an analysis are given in [16], [17], [18].

## Acknowledgment

We are thankful to Vladimir Bolkhovsky and Ravi Rastogi for overseen the wafer fabrication runs of the SFQ5ee and SFQ5ee+ processes used in this work.